# Analysis of unidirectional non-paraxial invisibility of purely reflective *PT*-symmetric volume gratings

Mykola Kulishov,[1*] H.F. Jones,[2] and Bernard Kress[3]

[1]*HTA Photomask, 1605 Remuda Lane, San Jose, California 95112, USA*
[2] *Physics Department, Imperial College, London, SW7 2BZ, UK*
[3]*Google, 1600 Amphitheatre Parkway, Mountain View, California 94043, USA*
[*]*mykolak@htaphotomask.com*

**Abstract:** We study the diffraction produced by a slab of purely reflective *PT*-symmetric volume Bragg grating that combines modulations of refractive index and gain/loss of the same periodicity with a quarter-period shift between them. Such a complex grating has a directional coupling between the different diffraction orders, which allows us to find an analytic solution for the first three orders of the full Maxwell equations without resorting to the paraxial approximation. This is important, because only with the full equations can the boundary conditions, allowing for the reflections, be properly implemented. Using our solution we analyze unidirectional invisibility of such a grating in a wide variety of configurations.

**References and links**

1. S. Bernet, S. B. Altner, F. R. Graf, E. S. Maniloff, A. Renn, and U. P. Wild, "Frequency and phase swept holograms in spectral hole-burning materials," Appl. Opt., **34**, 4674-4684 (1995).
2. L. Poladian, "Resonance mode expansions and exact solutions for nonuniform gratings," Phys. Rev. E., **54**, 2963-2975 (1996).
3. M. Kulishov, J. M. Laniel, N. Belanger, J. Azana, and D. V. Plant, "Nonreciprocal waveguide Bragg gratings," Optics Express, **13**, 3068–3078 (2005).
4. C. Keller, M. K. Oberthaler, R. Abfaiterer, S. Bernet, J. Schmiedmayer, and A. Zeilinger, "Tailored complex potentials and Friedel's law in atom optics," Phys. Rev. Lett., **79**, 3327-3330 (1997).
5. M. V. Berry, "Lop-sided diffraction by absorbing crystals," J. Phys. Math. Gen., **31**, 3493–3502 (1998).
6. C. M. Bender, and S. Boettcher, "Real spectra in non-Hermitian Hamiltonians having PT symmetry," Phys. Rev.Lett., **80**, 5243-5246 (1998).
7. C. M. Bender, S. Boettcher, and H. F. Jones, "Complex extension of quantum mechanics," Phys. Rev. Lett., **89**, 270401 (2002).
8. Z. Lin, H. Ramezani, T. Eichelkraut, T. Kottos, H. Cao, and D. Christodoulides, "Unidirectional invisibility induced by PT-symmetric periodic structures," Phys. Rev. Lett. **106**, 213901 (2011).
9. L. Feng, Y.-L. Xu, W. S. Fegadolli, M.-H. Lu, J. E. B. Olivera, V. R. Almeida, and Y.-F. Chen, A. Scherer, "Experimental demonstration of a unidirectional reflectionless parity-time metamaterial at optics frequencies," Nature Materials, **12**, 108-113 (2012).
10. Y. Yan, and N. C. Giebink, "Passive PT symmetry in organic composite films via complex refractive index modulation," Advanced Optical Materials, **2**, 423-427 (2014).
11. L. Feng, X. Zhu, S. Yang, H. Zhu, P. Zhang, X. Yin, Y. Wang, and X. Zhang, "Demonstration of a large-scale optical exceptional point structure," Optics Express, **22**, 1760-1767 (2014).
12. S. Longhi, "Invisibility in PT-symmetric complex crystals," J. Phys. A. **44**, 485-302, (2011).
13. X. Zhu, L. Feng, P. Zhang, X. Yin, and X. Zhang, "One way invisible cloak using parity-time symmetric transformation optics," Opt. Lett., **38**, 2821-2824 (2013).
14. C. Huang, F. Ye, Y.V. Kartashov, D.A. Malomed, and X. Chen, "PT-symmetry in optics beyond the paraxial approximation", Opt. Lett., **39**, 5443-5446 (2014).
15. T. K. Gaylord, and M. G. Moharam, "Planar dielectric grating diffraction theories," Appl. Phys. B, **28**, 1-14 (1982).
16. T. K. Gaylord, and M.G. Moharam, "Analysis and applications of optical diffraction by gratings," Proc. IEEE, **73**, 894-937 (1983).
17. M. Kulishov, H. F. Jones, and B. Kress, "Analysis of PT-symmetric volume gratings beyond the paraxial approximation," Optics Express, **23**, 9347-9362 (2015).

# 1. Introduction

Strongly asymmetric diffraction into conjugate diffraction orders was reported by S. Bernet et al. [1] as early as in 1995 in their experiments with transmission holograms made in hole-burning materials. The diffraction occurred on refractive index and absorption gratings recorded with a phase shift. In 1996 Poladian [2] applied modal analysis, well established for theoretical description of linear waveguides with a variety of cross-sections, to study linear gratings in waveguides. Using this analysis he demonstrated theoretically a nonreciprocal behavior when the grating profile combines index and gain/loss modulations. Such a structure has different reflection spectra when viewed from opposite ends. At the same time it was emphasized that the transmission through the structure must be *the same* from both sides, and in a strict sense such gratings do not satisfy the Lorentz reciprocity condition, so that devices such as optical isolators cannot be designed purely on the basis of such gratings. The proposed grating design was studied in detail [3] using Kogelnik's coupled wave theory for waveguide gratings with index and gain/loss modulation. Such a configuration is realized when the index and gain/loss gratings overlap in the waveguide, and have the same period, but are shifted by a half period in respect to one another: $\Delta \tilde{n} = a\cos(2\pi z/\Lambda) + jb\sin(2\pi z/\Lambda)$. Perfect reflection asymmetry is achieved when the gratings have the same amplitude ($a=b$). In such a case the complex perturbation profile becomes a pure phase: $\Delta \tilde{n} \Box \exp(2j\pi z/\Lambda)$. In this study [3] it was shown that the grating is not only transparent, as Poladian stated, but, in fact, light passing from one side of the grating does not change its amplitude or phase: such a behavior is classified as unidirectional invisibility.

On the other hand, such a potential attracted physicists working on the optics of atoms. Strong asymmetry of Bragg diffraction of atoms on a tailored complex potentials made of two overlapping standing light waves was experimentally demonstrated [4] in 1997, and in 1998 Berry [5] studied theoretically diffraction of atoms by a particular absorbing "crystal of light" described by the complex potential $\exp(2j\pi z/\Lambda) - 1$. Using the Raman-Nath diffraction equations he confirmed a strong diffraction asymmetry in transmission on such a complex potential which he called "lop-sided diffraction", in which diffraction occurs only in zeroth and positive (or only in zeroth and negative) diffraction orders.

About the same time Bender and Boettcher [6] relaxed one of the fundamental axioms of quantum mechanics regarding the Hermiticity of the Hamiltonian operator associated with real eigenvalues for Hamiltonians of any real quantum object. Replacing the Hermiticity condition by the new concept of Parity-Time (*PT*) symmetry, they demonstrated theoretically that entirely real eigenvalue spectra can exist for such non-Hermitian Hamiltonians [6, 7]. Although the impact of *PT*-symmetry in quantum mechanics is still discussed, its notions have been successfully extended and observed in optics by utilizing the isomorphism between the Schrödinger equation in quantum mechanics and the wave equation in the paraxial approximation [8]. It turned out that the grating profile proposed by Poladian [2] in 1996 can be perfectly characterized as a *PT*-symmetric structure, with its complex refractive index obeying $n(z) = n^*(-z)$, i.e. even/odd symmetry for real/imaginary parts of the refractive index.

As already mentioned, among many unusual and exotic characteristics of *PT*-symmetric structures, unidirectional invisibility is one of most intriguing, i.e. an input wave in a waveguide propagating from one side of the balanced (between index and gain/loss modulations) grating does not change either its amplitude or phase, which renders such a grating (or *PT*-metamaterial) completely invisible from that side. The general concept was confirmed experimentally [1, 4, 9, 10, 11] using a *PT*-symmetric passive structure without gain, i.e., with combined modulation of index and loss. Such a structure is reflectionless from one of its side; however, the wave attenuation does not allow us to call it *invisible*.

More rigorous analysis of invisibility in reflective, one-dimensional gratings by Longhi [12] has shown that for a given index and gain/loss modulation depth, three regimes are encountered as the grating length is increased. At short lengths the *PT*-symmetric grating is reflectionless and, when a signal is launched from one side, it demonstrates *unidirectional invisibility*. At intermediate lengths the grating remains reflectionless but not invisible; and finally, for even longer gratings both unidirectional reflectionless and invisibility are lost. In the current paper we will stay within the first of these regimes.

The phenomenon of invisibility has been studied mostly for reflective *PT*-symmetric gratings in waveguides in the paraxial approximation, whereas many applications, including invisibility cloaking, require *PT*-symmetric (grating) metamaterials in a slab with clearly defined boundaries. For example, many studies of invisibility were focused on perfect cloaks, to make the concealed object omnidirectionally invisible to outside observers. However, in many applications it might be important to have the cloak selectively visible for only one direction, i.e. a one-way cloak [13]. Such applications require knowledge of the influence of the slab boundary on the optical characteristics of the *PT*-metamaterial in reflection and transmission for non-paraxial light propagation. The only other study we are aware of that focused on non-paraxial behavior of the *PT*-symmetric structures [14] was concerned with the breaking and restoration of *PT*-symmetry in subwavelength guided structures and pure transmission gratings.

Traditionally diffractive properties were analyzed on the basis of coupled wave differential equations in which second-order derivatives were neglected. Such an approach is justified for one-dimensional gratings in optical waveguides, where the gratings represent weak modulation of the refractive index (its real and/or imaginary part) without any significant changes in its average value in the grating portion of the waveguide. In the case of slab gratings, illustrated in Fig. 1, neglecting the second derivatives of the field amplitudes is equivalent to neglecting the boundary effects for electromagnetic waves, i.e. the bulk diffracted orders are retained while the waves produced at the boundaries are eliminated. Such an approximation could lead to significant errors in some situations. In the case of *PT*-symmetric gratings, where the diffraction modes have a very unusual interaction mechanism, it is important to study how the slab boundaries affect the diffraction and how they affect invisibility in *PT*-symmetric volume grating.

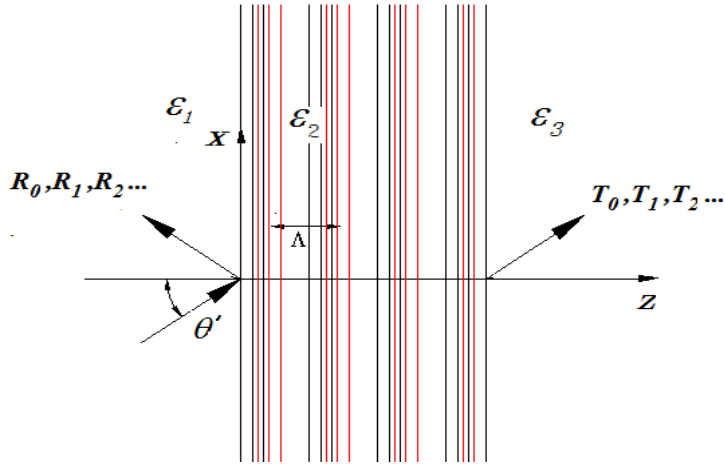

Fig. 1. Planar purely reflective grating of the index (black color fringes) and gain/loss (red color fringes) modulation.

We have therefore analyzed diffraction from such a slab by using the full, second-order Maxwell equations. Due to the particular directed structure of the coupled equations we are able to derive analytic expressions for the first three diffractive orders, In the following Sections 3 and 4 we use these expressions to analyze the properties of the *PT*-grating in a variety of different configurations characterized by the values of the background diffractive index within and on either side of the slab. A discussion of the general properties of this type of grating along with our conclusions is given in Sec. 5.

## 2. Analytic solution for first three Bragg orders for a balanced *PT*-symmetric grating

In this paper we study the diffraction characteristics of active holographic gratings as a gain/loss modulation in combination with traditional index gratings. The purely reflective grating is assumed to be composed of modulation of the relative dielectric permittivity

$$\varepsilon(x,z) = \varepsilon_2 + \Delta\varepsilon \cos(Kz) \qquad (1)$$

and modulation of gain and loss

$$\sigma(x,z) = \Delta\sigma \sin(Kz) \qquad (2)$$

in the region from $z = 0$ to $z = d$ with the same spatial frequency shifted by a quarter of period $\Lambda/4$ ($K=2\pi/\Lambda$) with respect to one another, where $\varepsilon_2$ is the average relative permittivity in the grating area; $\Delta\varepsilon$ is the amplitude of the sinusoidal relative permittivity, and $\Delta\sigma$ is the amplitude of the gain/loss periodic distribution. Unlike traditional modulation of the refractive index, Eq. (2) describes modulation of its imaginary part, so we will call the grating of Eq. (1) the real grating, and the grating described by Eq. (2) the imaginary one. Fig. 1 shows the generalized model of the hologram grating used in our study. It covers the case of free-space to free-space diffraction as well as planar slab holograms. The propagation constant $k(x,z)$ inside the grating slab is spatially modulated and related to the relative permittivity $\varepsilon(x,z)$ and gain/loss distribution $\sigma(x,z)$ by the well-known formula:

$$k^2(x,z) = k_0^2 \varepsilon(x,z) - j\omega\mu\sigma(x,z) \qquad (3)$$

where $\mu$ is the permeability of the medium, $\omega$ is the angular frequency of the wave and $k_0 = \omega/c$ is the wave-vector in free space, related to the free-space wavelength $\lambda_0$ by $k_0 = 2\pi/\lambda_0$.

Equations (1) - (3) can be combined in the following form:

$$k^2(x,z) = k_2^2 + 2k_2\kappa^- \exp(j\vec{K}\vec{r}) + 2k_2\kappa^+ \exp(-j\vec{K}\vec{r}) \qquad (4)$$

where $k_2 = k_0(\varepsilon_2)^{1/2}$ is the average propagation constant and $\vec{r}$ is the coordinate vector. The coupling constants $\kappa^+$ and $\kappa^-$ are

$$\kappa^\pm = \frac{1}{4(\varepsilon_2)^{1/2}}\left(k_0\Delta\varepsilon \pm c\mu\Delta\sigma\right) \qquad (5)$$

They can take quite different values, unlike the situation with only real or imaginary gratings where the coupling constants are always equal, at least in magnitude.

In the two unmodulated regions, $z < 0$ and $z > d$, where we assume uniform permittivity $\varepsilon_1$ and $\varepsilon_3$, respectively, the assumed solutions of the wave equation for the normalized electric fields are, for $z<0$ (incident wave and the reflected waves):

$$E_1(x,z) = \exp(-jk_1(x\sin(\theta') + z\cos(\theta'))) + \sum_{m=-\infty}^{m=+\infty} R_m \exp[-j\{k_2 x\sin\theta - [k_1^2 - k_2^2 \sin^2\theta]^{1/2} z\}] \qquad (6)$$

and for $z> 0$ (transmitted waves):

$$E_3(x,z) = \sum_{m=-\infty}^{m=+\infty} T_m \exp[-j\{k_2 x\sin\theta + [k_3^2 - k_2^2 \sin^2\theta]^{1/2}(z-d)\}] \qquad (7)$$

The total electric field in the hologram region (0< z< d) is the superposition of multiple waves:

$$E_2(x,z) = \sum_{m=-\infty}^{m=+\infty} S_m(z)\exp(-jk_2 x\sin\theta) \qquad (8)$$

where $k_1 = k_0(\varepsilon_1)^{1/2}$; $k_3 = k_0(\varepsilon_3)^{1/2}$; $\theta'$ is the angle of incidence in Region 1, and $\theta$ is the angle of refraction in Region 2, related to each other by $k_1 \sin\theta' = k_2 \sin\theta$. In these equations $R_m$, and $T_m$ are the amplitudes of the *m*-th reflected and *m*-th transmitted waves and are to be determined. $S_m(z)$ is the amplitude of the *m*-th wave anywhere in the modulated region and is to be determined by solving the wave equation for an incident plane wave with TE polarization (i.e. electric field perpendicular to the plane of incidence)

$$\nabla^2 E_2(x,z) + k_0^2 \varepsilon(x,z) E_2(x,z) = 0 \qquad (9)$$

To find $S_m(z)$, Eqs. (1) and (8) are substituted into Eq. (9), resulting in the system of coupled-wave equations [14]:

$$\frac{d^2 S_m(z)}{dz^2} - k_2^2 \cos^2\theta S_m(z) + 2k_2\left[\kappa^- \exp(jKz)S_{m+1}(z) + \kappa^+ \exp(-jKz)S_{m-1}(z)\right] = 0 \qquad (10)$$

This set of coupled-wave equations contains no first-derivative terms. In addition, Eqs. (10) are nonconstant-coefficient differential equations due to the presence of *z* in the coefficients of the $S_{m-1}(z)$ and $S_{m+1}(z)$ terms. Normally this system of equations is used only for purely transmission diffraction gratings, which eliminates the coordinate dependence in the $S_{m-1}(z)$ and $S_{m+1}(z)$ terms. However, in the case of a perfectly balanced *PT*-symmetric grating one of the coupling coefficients is zero. In this paper we will deal with purely reflective *PT*-symmetric gratings, leaving the general case to a further publication.

Pure reflection dielectric gratings with fringes parallel to the surface ($\phi = 0$, where $\phi$ is the angle that the fringe planes makes with the *z* axis, as shown in Fig. 1) have been analyzed using rigorous coupled-wave analysis [15, 16]. There is a significant distinction between such pure reflection gratings and the slanted-fringe reflection gratings, with $\phi \neq 0$. The "grating equation" for the backward-diffracted waves: $\sqrt{\varepsilon_1}\sin\theta'_m = \sqrt{\varepsilon_1}\sin\theta' - m(\lambda/\Lambda)\sin\phi$, demonstrates that for $\phi = 0$ there is a continuum of solutions that contribute in the same angle of diffraction. However, for $\phi \neq 0$ the continuum of solutions disappears, and each diffraction order *m* has its specific diffraction angle. In the limit of unslanted fringes ($\phi = 0$), the field inside the grating is not periodic along the boundary. For any value of *z* inside the grating, the field is composed of two plane waves, one with a component in the positive *z* direction and one with a component in the negative *z* direction. This field is phase matched to only two waves outside the grating. These are the transmitted wave *(T)* and the reflected wave *(R),* as shown in Fig. 1. In this limit, for left-side incidence the coupled wave equations (10) for the dimensionless coordinate $u = k_2 z$ reduce to the following form:

$$\frac{d^2 S_m(z)}{du^2} + \cos^2\theta S_m(z) + \xi\exp(j\gamma u)S_{m-1}(z) = 0 \qquad (11)$$

where $\gamma = \lambda/(\Lambda(\varepsilon_2)^{1/2}) = 2\cos\theta_B$, with $\theta_B = \arccos(\lambda/(2\Lambda(\varepsilon_2)^{1/2})$, and they become a system of inhomogeneous second order differential linear equations that can be solved analytically. Here $\theta_B$ is the first-order Bragg diffraction angle. It can be seen from this equation that the coupling occurs unidirectionally, from zeroth order to the positive first (+1) order, from the positive first order to the positive second (+2) order and so on. There is no coupling from zeroth order into the first negative (-1) and all higher negative orders: $S_{-m} = 0$ ($m = 1, 2, 3…$).

The *PT*-symmetric reflection gratings are not symmetrical in the *z*-direction. The complex profile: $\Delta\tilde{n} = \cos(2\pi z/\Lambda) + j\sin(2\pi z/\Lambda) = \exp(2j\pi z/\Lambda)$ for a light wave incident from the left side will be "viewed" as $\Delta\tilde{n} = \cos(2\pi z/\Lambda) - j\sin(2\pi z/\Lambda) = \exp(-2j\pi z/\Lambda)$ for a light wave incident on the slab from the right side. For right-side incidence the coupling coefficient $\kappa^- = 0$ in Eq. (10) will be zero for the balanced PT-symmetric grating and $\kappa^+ \neq 0$, and Eq. (10) then becomes:

$$\frac{d^2 S_m(u)}{du^2} + \cos^2\theta S_m(u) + \xi\exp(-j\gamma u) S_{m+1}(u) = 0 \qquad (12)$$

In this situation the PT-symmetric grating provides power transfer from zeroth order into the first negative diffraction order ($m = -1$), from the first negative order to the second negative order ($m = -2$) and so on. Therefore, for light incidence from the right side diffraction takes place into the negative orders, and there is no diffraction into positive orders ($S_m = 0$ for $m \geq 1$).

These equations (11) and (12) require the boundary conditions that the tangential electric and tangential magnetic fields be continuous across the two boundaries ($z = 0$ and $z = d$). For the H-mode polarization described in this paper, the electric field only has a component in the y-direction, and so it is the tangential electric field directly. The magnetic field intensity, however, must be obtained through the Maxwell equation. The tangential component of H is in the x-direction and is thus given by $H_x = (-j/\omega\mu_0)\partial E_y/\partial z$. The resulting boundary conditions are:

a) tangential E at $z = 0$:

$$\delta_{0m} + R_m(0) = S_m(0); \qquad (13)$$

b) tangential H at $z = 0$:

$$\frac{dS_m(0)}{du}\frac{du}{dz} = j(k_1^2 - k_2^2\sin^2\theta)^{1/2}(R_m - \delta_{0m}); \qquad (14)$$

c) tangential E at $z = d$:

$$T_m(d) = S_m(d) \qquad (15)$$

d) tangential H at $z = d$:

$$\frac{dS_m(d)}{du}\frac{du}{dz} = -j(k_3^2 - k_2^2\sin^2\theta)^{1/2}T_m. \qquad (16)$$

*2.1 Zeroth diffractive orders in transmission and reflection*

For the zeroth-order amplitude $S_0(u)$ (non-diffracted light) Eqs. (11) and (12) are identical and are decoupled from the second equation for the first-order amplitudes, as in the case of the purely transmission grating [16] $\phi=\pi/2$. We therefore have an equation for $S_0(u)$ of the form:

$$\frac{d^2 S_0(u)}{du^2} + \eta_0^2 S_0(u) = 0 \qquad (17)$$

with solution

$$S_0(u) = A_0\exp(ju\eta_0) + B_0\exp(-ju\eta_0) \qquad (18)$$

where the eigenvalue is $\eta_0 = \cos\theta$. Applying the boundary conditions (13)-(16) we can find $T_0$ and $R_0$ and the constants A and B:

$$T_0 = \frac{4\alpha_0\eta_0}{(\alpha_0+\eta_0)(\beta_0+\eta_0)\exp(ju_d\eta_0) - (\alpha_0-\eta_0)(\beta_0-\eta_0)\exp(-ju_d\eta_0)} \qquad (19)$$

$$R_0 = \frac{(\eta_0+\alpha_0)(\eta_0-\beta_0)\exp(-ju_d\eta_0) - (\eta_0-\alpha_0)(\eta_0+\beta_0)\exp(ju_d\eta_0)}{(\eta_0+\alpha_0)(\eta_0+\beta_0)\exp(ju_d\eta_0) - (\eta_0-\alpha_0)(\eta_0-\beta_0)\exp(-ju_d\eta_0)} \qquad (20)$$

$$A_0 = \frac{T_0}{2}\left(\frac{\eta_0 - \beta_0}{\beta_0}\right)\exp(-ju_d\eta_0) \qquad B_0 = \frac{T_0}{2}\left(\frac{\eta_0 + \beta_0}{\beta_0}\right)\exp(ju_d\eta_0) \qquad (21)$$

where $\alpha_0 = \sqrt{\varepsilon_1/\varepsilon_2 - \sin^2\theta}$; $\beta_0 = \sqrt{\varepsilon_3/\varepsilon_2 - \sin^2\theta}$.

We can see from these equations that when $\varepsilon_3 = \varepsilon_2 = \varepsilon_1$ (no reflection from the interfaces between Region 1 and 2 and Regions 2 and 3), then $\alpha_0 = \beta_0 = \eta_0 = \cos\theta$, so that $T_0 = \exp(-jk_2 d\cos\theta)$, and $R_0 = 0$, i.e. zeroth order (non-diffractive portion of the incident light) propagates through the hologram without any attenuation or amplification practically unaffected by the *PT*-symmetric volume grating.

## 2.1 Higher diffractive orders

Equations (11) and (12) can be reduced to a system of nonhomogeneous second-order differential equations, whose solutions are sought as a sum of the general solution of the homogenous second-order differential equations and the particular solution of the inhomogeneous second-order differential equations. The characteristic equations for the homogenous second-order differential equations have the same eigenvalue: $\eta_0 = \cos\theta$. Therefore the solution of the homogeneous equation for any *m*-th diffractive order will be of the same form: $A\exp(+j\eta_0 u) + B\exp(-j\eta_0 u)$, where *A* and *B* are to be found from the boundary conditions for each order.

## 2.2.1. First diffraction orders

For example, for the first-order diffraction the differential equation for left-side incidence is:

$$\frac{d^2 S_1(u)}{du^2} + \eta_0^2 S_1(u) + \xi\exp(j\gamma u) S_0(u) = 0 \qquad (22)$$

and for right-side incidence

$$\frac{d^2 S_{-1}(u)}{du^2} + \eta_0^2 S_{-1}(u) + \xi\exp(-j\gamma u) S_0(u) = 0 \qquad (23)$$

The solutions, $S_{\pm 1}(u)$, can be found as a sum of the general solution of the homogenous equation, $(S_{\pm 1})_H$, and a particular solution $(S_{\pm 1})_P$ of the nonhomogeneous equation. The solution of the homogeneous equation is:

$$(S_{\pm 1}(u))_H = A_{\pm 1}\exp(j\eta_0 u) + B_{\pm 1}\exp(-j\eta_0 u) \qquad (24)$$

The particular solution can be found using the method of undetermined coefficients. We write

$$(S_{\pm 1}(u))_P = \xi A_0 X_0^{(\pm)}\exp(j\eta_0 u) + \xi B_0 Y_0^{(\pm)}\exp(j\eta_0 u) \qquad (25)$$

where

$$X_0^{(\pm)} = \frac{1}{\gamma(2\eta_0 \pm \gamma)} \qquad Y_0^{(\pm)} = \frac{1}{\gamma(\gamma \mp 2\eta_0)} \qquad (26)$$

Applying the boundary conditions (13)-(16) we can find $A_1$ and $B_1$ and $R_1$ and $T_1$, namely:

$$R_{\pm1} = \frac{A_0\xi\left[\frac{\eta_0+\beta_0}{\pm\gamma}(e^{j\eta_0 u_d}-e^{j(\eta_0\pm\gamma)u_d})-\frac{\eta_0-\beta_0}{2\eta_0\pm\gamma}(e^{j(\eta_0\pm\gamma)u_d}-e^{-j\eta_0 u_d})\right]}{(\eta_0+\alpha_0)(\eta_0+\beta_0)\exp(j\eta_0 u_d)-(\eta_0-\alpha_0)(\eta_0-\beta_0)\exp(-j\eta_0 u_d)} +$$
$$+\frac{B_0\xi\left[\frac{\eta_0+\beta_0}{\pm\gamma-2\eta_0}(e^{j\eta_0 u_d}-e^{j(\pm\gamma-\eta_0)u_d})-\frac{\eta_0-\beta_0}{\pm\gamma}(e^{j(\pm\gamma-\eta_0)u_d}-e^{-j\eta_0 u_d})\right]}{(\eta_0+\alpha_0)(\eta_0+\beta_0)\exp(j\eta_0 u_d)-(\eta_0-\alpha_0)(\eta_0-\beta_0)\exp(-j\eta_0 u_d)}$$

(27)

$$T_{\pm1} = \frac{A_0\xi\left[\frac{\eta_0-\alpha_0}{\pm\gamma}(1-e^{\pm j\gamma u_d})+\frac{\eta_0+\alpha_0}{2\eta_0\pm\gamma}(1-e^{j(2\eta_0\pm\gamma)u_d})\right]}{(\eta_0+\alpha_0)(\eta_0+\beta_0)\exp(j\eta_0 u_d)-(\eta_0-\alpha_0)(\eta_0-\beta_0)\exp(-j\eta_0 u_d)} +$$
$$+\frac{B_0\xi\left[\frac{\eta_0-\alpha_0}{\pm\gamma-2\eta_0}(1-e^{-j(2\eta_0\mp\gamma)u_d})+\frac{\eta_0+\alpha_0}{\pm\gamma}(1-e^{\pm j\gamma u_d})\right]}{(\eta_0+\alpha_0)(\eta_0+\beta_0)\exp(j\eta_0 u_d)-(\eta_0-\alpha_0)(\eta_0-\beta_0)\exp(-j\eta_0 u_d)}$$

(28)

$$A_{\pm1} = \frac{T_{\pm1}}{2}\left(\frac{\eta_0-\beta_0}{\eta_0}\right)e^{-j\eta_0 u_d} \pm \frac{\xi A_0}{2\gamma\eta_0}e^{\pm j\gamma u_d} - \frac{\xi B_0}{2\eta_0(\pm\gamma-2\eta_0)}e^{j(\pm\gamma-\eta_0)u_d} \quad (29)$$

$$B_{\pm1} = \frac{T_{\pm1}}{2}\left(\frac{\eta_0+\beta_0}{\eta_0}\right)e^{+j\eta_0 u_d} + \frac{\xi A_0}{2\eta_0(\pm\gamma+2\eta_0)}e^{j(\pm\gamma+\eta_0)u_d} \pm \frac{\xi B_0}{2\gamma\eta_0}e^{\pm j\gamma u_d} \quad (30)$$

The mode coupling in a *PT*-symmetric grating in its balanced state has a unidirectional nature, which makes it relatively easy to find practically any higher diffraction order analytically. Here we exploit this feature to derive explicit expressions for the second-order reflection and transmission coefficients.

*2.2.2. Second diffraction orders*

We need to solve the following differential equation to obtain the second diffraction orders $R_{\pm2}$ and $T_{\pm2}$

$$\frac{d^2 S_{\pm2}(u)}{du^2} + \eta_0^2 S_{\pm2}(u) + \xi\exp(\pm j\gamma u) S_{\pm1}(u) = 0 \quad (31)$$

Similar to the solution of Eq. (23), $S_{\pm2}(u)$, can be found as a sum of the general solution of the homogenous equation, $(S_{\pm2})_H$ and a particular solution $(S_{\pm2})_P$ of the inhomogeneous equation.

$$(S_{\pm2}(u))_H = A_{\pm2}\exp(j\eta_0 u) + B_{\pm2}\exp(-j\eta_0 u) \quad (32)$$

$$(S_{\pm2}(u))_P = A_{\pm2}\exp(j(\eta_0\pm\gamma)u) + B_{\pm2}\exp(j(-\eta_0\pm\gamma)u) +$$
$$+ X_1^{(\pm)}\exp(j(\eta_0\pm 2\gamma)u) + Y_1^{(\pm)}\exp(j(-\eta_0\pm 2\gamma)u) \quad (33)$$

where

$$A_{\pm2} = \frac{\xi A_{\pm1}}{\pm\gamma(\pm\gamma+\eta_0)} \qquad B_{\pm2} = \frac{\xi B_{\pm1}}{\pm\gamma(\pm\gamma-\eta_0)} \quad (34)$$

$$X_1^{(\pm)} = \frac{\xi^2 A_0}{4\gamma^2(\pm\gamma+2\eta_0)(\pm\gamma+\eta_0)}; \qquad Y_1^{(\pm)} = \frac{\xi^2 B_0}{4\gamma^2(\pm\gamma-2\eta_0)(\pm\gamma-\eta_0)}; \quad (35)$$

Again, applying the boundary conditions (13)-(16) we can find $A_{\pm2}$ and $B_{\pm2}$ and $R_{\pm2}$ and $T_{\pm2}$:

$$R_{\pm 2} = \frac{A_{\pm 2}\left[(\pm\gamma+2\eta_0)(\eta_0+\beta_0)(e^{j\eta_0 u_d}-e^{j(\pm\gamma+\eta_0)u_d})\pm\gamma(\eta_0-\beta_0)(e^{-j\eta_0 u_d}-e^{j(\pm\gamma+\eta_0)u_d})\right]}{(\eta_0+\alpha_0)(\eta_0+\beta_0)\exp(j\eta_0 u_d)-(\eta_0-\alpha_0)(\eta_0-\beta_0)\exp(-j\eta_0 u_d)}+$$

$$+\frac{B_{\pm 2}\left[\pm\gamma(\eta_0+\beta_0)(e^{j\eta_0 u_d}-e^{j(\pm\gamma-\eta_0)u_d})+(\pm\gamma-2\eta_0)(\eta_0-\beta_0)(e^{-j\eta_0 u_d}-e^{j(\pm\gamma-\eta_0)u_d})\right]}{(\eta_0+\alpha_0)(\eta_0+\beta_0)\exp(j\eta_0 u_d)-(\eta_0-\alpha_0)(\eta_0-\beta_0)\exp(-j\eta_0 u_d)}+$$

$$+\frac{2X_1^{(\pm)}\left[(\pm\gamma+\eta_0)(\eta_0+\beta_0)(e^{j\eta_0 u_d}-e^{j(\pm 2\gamma+\eta_0)u_d})\pm\gamma(\eta_0-\beta_0)(e^{-j\eta_0 u_d}-e^{j(\pm 2\gamma+\eta_0)u_d})\right]}{(\eta_0+\alpha_0)(\eta_0+\beta_0)\exp(j\eta_0 u_d)-(\eta_0-\alpha_0)(\eta_0-\beta_0)\exp(-j\eta_0 u_d)}+$$

$$+\frac{2Y_1^{(\pm)}\left[\pm\gamma(\eta_0+\beta_0)(e^{j\eta_0 u_d}-e^{j(\pm 2\gamma-\eta_0)u_d})+(\pm\gamma-\eta_0)(\eta_0-\beta_0)(e^{-j\eta_0 u_d}-e^{j(\pm\gamma-\eta_0)u_d})\right]}{(\eta_0+\alpha_0)(\eta_0+\beta_0)\exp(j\eta_0 u_d)-(\eta_0-\alpha_0)(\eta_0-\beta_0)\exp(-j\eta_0 u_d)}$$

(36)

$$T_{\pm 2} = \frac{A_{\pm 2}\left[(\pm\gamma+2\eta_0)(\eta_0-\alpha_0)(1-e^{\pm j\gamma u_d})\pm\gamma(\eta_0+\alpha_0)(1-e^{j(\pm\gamma+2\eta_0)u_d})\right]}{(\eta_0+\alpha_0)(\eta_0+\beta_0)\exp(j\eta_0 u_d)-(\eta_0-\alpha_0)(\eta_0-\beta_0)\exp(-j\eta_0 u_d)}+$$

$$+\frac{B_{\pm 2}\left[\pm\gamma(\eta_0-\alpha_0)(1-e^{j(\pm\gamma-2\eta_0)u_d})+(\pm\gamma-2\eta_0)(\eta_0+\alpha_0)(1-e^{\pm j\gamma u_d})\right]}{(\eta_0+\alpha_0)(\eta_0+\beta_0)\exp(j\eta_0 u_d)-(\eta_0-\alpha_0)(\eta_0-\beta_0)\exp(-j\eta_0 u_d)}+$$

$$+\frac{2X_1^{(\pm)}\left[(\pm\gamma+\eta_0)(\eta_0-\alpha_0)(1-e^{\pm j2\gamma u_d})\pm\gamma(\eta_0+\alpha_0)(1-e^{j2(\pm\gamma+\eta_0)u_d})\right]}{(\eta_0+\alpha_0)(\eta_0+\beta_0)\exp(j\eta_0 u_d)-(\eta_0-\alpha_0)(\eta_0-\beta_0)\exp(-j\eta_0 u_d)}+$$

$$+\frac{2Y_1^{(\pm)}\left[\pm\gamma(\eta_0-\alpha_0)(1-e^{j2(\pm\gamma-\eta_0)u_d})+(\pm\gamma-\eta_0)(\eta_0+\alpha_0)(1-e^{\pm j2\gamma u_d})\right]}{(\eta_0+\alpha_0)(\eta_0+\beta_0)\exp(j\eta_0 u_d)-(\eta_0-\alpha_0)(\eta_0-\beta_0)\exp(-j\eta_0 u_d)}$$

(37)

We limit our analysis to the first and second diffraction orders, and therefore the expressions for $A_{\pm 2}$ and $B_{\pm 2}$ are not needed.

## 3. Filled-space *PT*-symmetric gratings: $\varepsilon_1 = \varepsilon_2 = \varepsilon_3$

This configuration exhibits the ideal unidirectional characteristics that such *PT*-symmetric gratings could exhibit. They are easy to study before we include into the analysis the reflection from the slab boundaries, which will degrade these ideal characteristics. Indeed, for such a *PT*-symmetric grating the main factors simplify to the following forms: $\alpha_0 = \alpha_1 = \alpha_2 = \beta_0 = \beta_1 = \beta_2 = \cos\theta$ ; $A_0=0$; $B_0=1$ and Eqs. (27), (28) and (36), (37) reduce to

$$R_1 = -j\frac{\xi}{2}\frac{\sin[(\cos\theta_B-\cos\theta)u_d]}{\cos\theta(\cos\theta_B-\cos\theta)}\exp(j(\cos\theta_B-\cos\theta)u_d) \quad (38)$$

$$R_{-1} = -j\frac{\xi}{2}\frac{\sin[(\cos\theta_B+\cos\theta)u_d]}{\cos\theta(\cos\theta_B+\cos\theta)}\exp(-j(\cos\theta_B+\cos\theta)u_d) \quad (39)$$

$$T_1 = -\frac{\xi}{4}\left[\frac{\exp(2j\cos\theta_B u_d)-1}{\cos\theta\cos\theta_B}\right]\exp(-ju_d\cos\theta) \quad (40)$$

$$T_{-1} = \frac{\xi}{4}\left[\frac{\exp(-2j\cos\theta_B u_d)-1}{\cos\theta\cos\theta_B}\right]\exp(-ju_d\cos\theta) \quad (41)$$

$$R_2 = j\frac{\xi^2\exp(j(\cos\theta_B-\cos\theta)u_d)}{4\cos^2\theta\cos\theta_B}\left[\frac{\sin[(2\cos\theta_B-\cos\theta)u_d]}{2\cos\theta_B-\cos\theta}\exp(j\cos\theta_B u_d)-\frac{\sin[(\cos\theta_B-\cos\theta)u_d]}{\cos\theta_B-\cos\theta}\right] \quad (42)$$

$$R_{-2} = -j\frac{\xi^2\exp(-j(\cos\theta_B+\cos\theta)u_d)}{4\cos^2\theta\cos\theta_B}\left[\frac{\sin[(2\cos\theta_B+\cos\theta)u_d]}{2\cos\theta_B+\cos\theta}\exp(-j\cos\theta_B u_d)-\frac{\sin[(\cos\theta_B+\cos\theta)u_d]}{\cos\theta_B+\cos\theta}\right] \quad (43)$$

$$T_2 = -j\frac{\xi^2 \exp(j(\cos\theta_B - \cos\theta)u_d)}{8\cos^2\theta}\left[\frac{\sin[(\cos\theta_B - \cos\theta)u_d]}{\cos^2\theta_B - \cos^2\theta}\exp(-j\cos\theta u_d) + \right.$$
$$\left. +\frac{\sin(\cos\theta_B u_d)}{\cos^2\theta_B} - \frac{\sin(2\cos\theta_B u_d)}{2\cos^2\theta_B}\frac{(2\cos\theta_B + \cos\theta)}{(\cos\theta_B + \cos\theta)}\exp(j\cos\theta_B u_d)\right] \quad (44)$$

$$T_{-2} = j\frac{\xi^2 \exp(-j(\cos\theta_B + \cos\theta)u_d)}{8\cos^2\theta}\left[\frac{\sin[(\cos\theta_B + \cos\theta)u_d]}{\cos^2\theta_B - \cos^2\theta}\exp(-j\cos\theta u_d) + \right.$$
$$\left. +\frac{\sin(\cos\theta_B u_d)}{\cos^2\theta_B} - \frac{\sin(2\cos\theta_B u_d)}{2\cos^2\theta_B}\frac{(2\cos\theta_B - \cos\theta)}{(\cos\theta_B - \cos\theta)}\exp(-j\cos\theta_B u_d)\right] \quad (45)$$

Comparing the expressions for the transmission in the first positive $T_1$ order (diffraction in the $+z$ direction Eq. (40)) and negative $T_{-1}$ order (diffraction in the $-z$ direction Eq. (41)), we see that the transmission coefficients are nonzero and differ by a phase. However, the system is only truly *PT* symmetric if we have an integral number of grating periods in the grating thickness, i.e. $d = m\Lambda$, where $m$ is an integer. In that case $2u_d \cos\theta_B = m\pi$ and the phase factors $\exp(\pm 2ju_d \cos\theta_B)$ are equal to 1, so that $T_1 = T_{-1} = 0$.

In fact, the condition of an integral number of periods in the grating thickness significantly simplifies the expressions for all non-zeroth diffractive orders. For example, Eqs. (27), (28) and (36), (37) reduce to

$$R_{\pm 1} = \frac{j\xi \sin(\eta_0 u_d)}{(\eta_0 \pm \gamma/2)}\left[\frac{A_0(\eta_0 - \beta_0) - B_0(\eta_0 + \beta_0)}{(\eta_0 + \alpha_0)(\eta_0 + \beta_0)\exp(j\eta_0 u_d) - (\eta_0 - \alpha_0)(\eta_0 - \beta_0)\exp(-j\eta_0 u_d)}\right] \quad (46)$$

$$T_{\pm 1} = \xi\frac{(1 - e^{j2\eta_0 u_d})}{2\eta_0 \pm \gamma}\left[\frac{A_0(\eta_0 + \alpha_0) - B_0(\eta_0 - \alpha_0)}{(\eta_0 + \alpha_0)(\eta_0 + \beta_0)\exp(j\eta_0 u_d) - (\eta_0 - \alpha_0)(\eta_0 - \beta_0)\exp(-j\eta_0 u_d)}\right] \quad (47)$$

$$R_{\pm 2} = \pm\gamma\frac{\left[(B_{\pm 2} + 2Y_1^{(\pm)})(\eta_0 + \beta_0) - (A_{\pm 2} + 2X_1^{(\pm)})(\eta_0 - \beta_0)\right](e^{j\eta_0 u_d} - e^{-j\eta_0 u_d})}{(\eta_0 + \alpha_0)(\eta_0 + \beta_0)\exp(j\eta_0 u_d) - (\eta_0 - \alpha_0)(\eta_0 - \beta_0)\exp(-j\eta_0 u_d)} \quad (48)$$

$$T_{\pm 2} = \pm\gamma\frac{(A_{\pm 2} + 2X_1^{(\pm)})(\eta_0 + \alpha_0)(1 - e^{j2\eta_0 u_d}) + (B_{\pm 2} + 2Y_1^{(\pm)})(\eta_0 - \alpha_0)(1 - e^{-j2\eta_0 u_d})}{(\eta_0 + \alpha_0)(\eta_0 + \beta_0)\exp(j\eta_0 u_d) - (\eta_0 - \alpha_0)(\eta_0 - \beta_0)\exp(-j\eta_0 u_d)} \quad (49)$$

Looking at Eq. (39) we can see right away that the resonance takes place only in reflection at $\theta = \pm\theta_B$ (Bragg angles) and only in positive diffraction orders, and the diffraction amplitudes reach the following values at the resonance:

$$R_1 = -j\frac{\xi u_d}{2\cos\theta_B}; \qquad T_1 = -j\frac{\xi \sin(u_d \cos\theta_B)}{2\cos^2\theta_B} \quad (50)$$

$$R_{-1} = -j\frac{\xi}{4}\frac{\sin(2\cos\theta_B u_d)}{\cos^2\theta_B}\exp(-2j\cos\theta_B u_d); \qquad T_{-1} = -j\frac{\xi}{2}\frac{\sin(\cos\theta_B u_d)}{\cos^2\theta_B}\exp(-2j\cos\theta_B u_d) \quad (51)$$

$$R_2 = j\frac{\xi^2}{4\cos^4\theta_B}\left[\sin(\cos\theta_B u_d)\exp(j\cos\theta_B u_d) - u_d\right] \quad (52)$$

$$R_{-2} = -j\frac{\xi^2 \exp(-2j\cos\theta_B u_d)}{4\cos^4\theta_B}\left[\frac{1}{3}\sin(3\cos\theta_B u_d)\exp(-j\cos\theta_B u_d) - \frac{1}{2}\sin(3\cos\theta_B u_d)\right] \quad (53)$$

$$T_2 = -j\frac{\xi^2}{8\cos^3\theta_B}\left[\frac{u_d}{2}\exp(-j\cos\theta_B u_d) + \frac{\sin(\cos\theta_B u_d)}{\cos\theta_B} - \frac{3}{4}\frac{\sin(2\cos\theta_B u_d)}{\cos\theta_B}\exp(j\cos\theta_B u_d)\right] \quad (54)$$

The purely reflective grating diffracts backward with an amplitude proportional to the grating strength, the product of the grating length and the coupling coefficient, $\xi u_d$. At the same time diffraction in transmission from the left side as well as diffraction from the right side is generally small but not exactly zero. However, for the *PT*-symmetric configuration ($d = m\Lambda$), $R_0 = T_1 = T_{-1} = R_{-1} = 0$, $T_0 = 1$, and $R_1 \neq 0$, i.e., we have the perfect case of grating invisibility from the right side for any angle of incidence. A typical example of diffraction from the left and right sides is shown in Fig. 2.

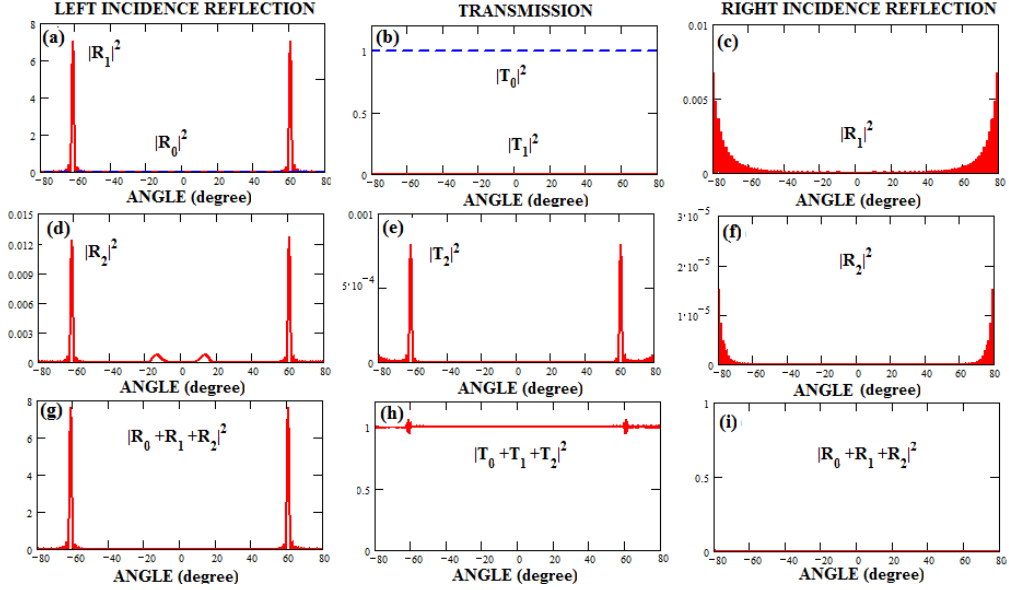

Fig. 2. Filled-space configuration ($\varepsilon_1 = \varepsilon_2 = \varepsilon_2 = 2.4$): (left side) zeroth (blue, dashed), first order (red, solid) (a) and second (d) order reflections, as well as the combined coherent reflection (g) and zeroth, first (b) and second (e) orders and the combined coherent transmission (h) along with (right side) zeroth, first order (c), second order (f) and the combined coherent reflection (i) as a function of the internal angle of incidence, $\theta$, for $\Lambda$=0.42 μm (red, solid). The other parameters are $d$ = 8.4 μm, $\lambda_0$ = 0.633 μm, $\xi$=0.02.

As we can see, in the filled-space configuration there is no reflection of non-diffracted light (blue dashed line in Fig. 2 (a)), and zero diffraction in the first and practically negligible diffraction in the second transmission orders (red solid curves in Fig. 2 (b) and Fig. 2 (e). Practically all diffraction occurs in the first reflection order (red solid curve in Fig.2 (a)) at the first diffraction angle $\theta_B^{(1)} = 60.9^0$ for left-side incidence. The second order contributes mainly in the first diffraction peaks $\theta_B^{(1)} = 60.9^0$, as well as in the second diffraction angle $\theta_B^{(2)} = 13.46^0$, although the total contribution is very small. The transmission characteristics (Figs. 2 (b), (e) and (h)) are the same for the left- and right-side incidences. The strong asymmetry of the *PT*-symmetric grating is revealed in the very strong difference in reflection from the two sides of the grating slab. Indeed, the contributions from the first- and second reflection-orders (Fig. 2 (c) and (f)) are negligible. Fig. 2 (e), Fig. 2 (f) and Fig. 2 (i) represent coherently added contributions from zeroth, first and second orders in reflection from the left, transmission, and reflection from the right, respectively, demonstrating practically perfect grating invisibility when viewed from the right side; such a *PT*-grating performance also can be called "see through" diffraction. This *PT*-symmetric grating behavior is summarized in Fig. 3.

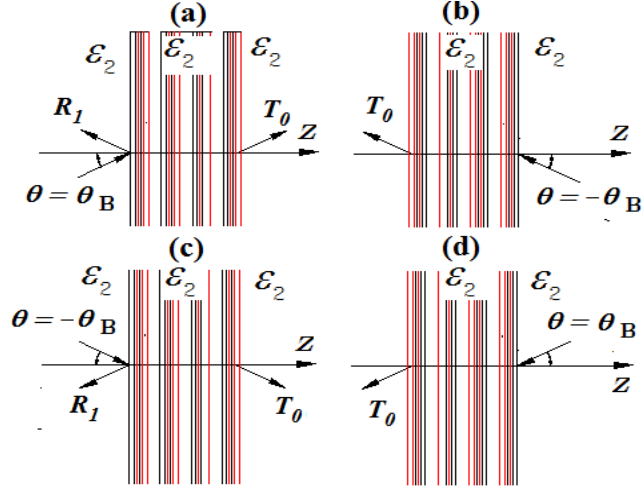

Fig. 3. Prominent modes of the PT-symmetric grating for incidence at different angles and from different sides: (a) from the left near the first Bragg angle $\theta_B$; (b) from the left near $-\theta_B$; (c) from the right near $\theta_B$; (d) from the right near $-\theta_B$.

Unlike the diffraction angle in a pure transmission grating [17], the diffraction angle in the pure reflection grating $\theta_B = \arccos(\lambda/(2\Lambda(\varepsilon_2)^{1/2}))$ gets smaller as the grating period ($\Lambda$) decreases. Fig. 4 shows two examples of diffraction in reflection (red, solid curves) and in transmission (blue dashed curves) and for two different values of $\Lambda$: $\Lambda=0.25$ μm (Fig. 4 (a)) and $\Lambda=0.205$ μm (Fig. 4 (b)). As we can see, at $\Lambda=0.25$ the diffraction occurs at $\theta_B^{(1)} = 35.22^0$ and for $\Lambda=0.205$ it takes place at $\theta_B^{(1)} = 4.95^0$, and the two peaks overlap (Fig. 4 (b)) forming a single reflection band with the center at $\theta = 0^0$. It is interesting that with decrease of the grating period and as the peaks move towards the center (the normal to the slab), they get wider and decrease in magnitude.

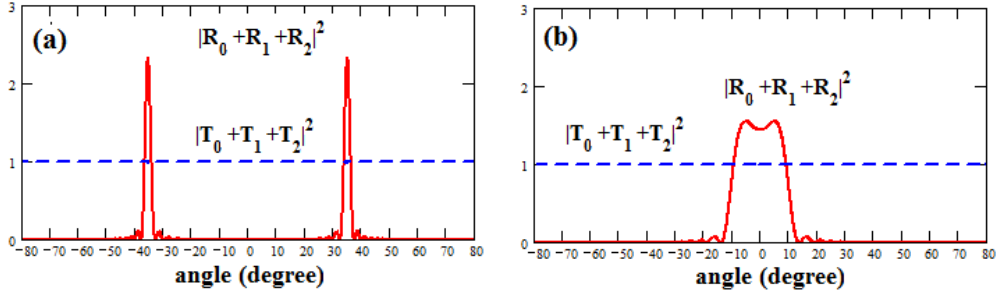

Fig. 4. Filled-space configuration ($\varepsilon_1 = \varepsilon_2 = \varepsilon_2 = 2.4$): the combined coherent zeroth, first and second orders in reflection (red solid curve) for left-side incidence and in transmission (blue, dashed curve) as functions of the internal angle of incidence $\theta$ for $\Lambda=0.25$ μm (a) $\Lambda=0.205$ (b). The other parameters are $d = 8$ μm, $\lambda_0 = 0.633$ μm, $\xi=0.02$.

## 4. Reflective *PT*-symmetric gratings with Fresnel reflections

*4.1. Symmetric geometry* $\varepsilon_1 = \varepsilon_3 = 1; \varepsilon_2 = 2.4$

The strongest interference of light reflected from the front and back interfaces of the slab takes place when there is a strong refractive index contrast between the slab (Region 2) and the surrounding media (Regions 1 and 3). It is achieved by placing the slab in air, i.e. $\varepsilon_1 = \varepsilon_3 = 1; \varepsilon_2 = 2.4$. Comparing Fig. 5 with Fig. 2 it is evident that this interference leads to a significant increase in the first-order diffraction in transmission $T_1$ (Fig. 5 (b) red curve) as well as the second-order diffraction in reflection (Fig. 5 (d)) and transmission Fig. 5 (e). It also strongly affects zeroth-order diffracted light both in reflection (Fig. 5 (a) blue) and in transmission (Fig. 5 (b) blue). For the purpose of consistency, all spectra are plotted against the internal incident angle $\theta$, which can be easily related to the external angle of incidence $\theta'$ and angle $\theta''$. The range of the angles is limited by $\pm 40^0$, which corresponds to external incident angles $\pm 90^0$. At the same time diffraction from the right side is relatively small (Fig. 5(c) red curve, Fig. 5(f)). The main distortions are due to Fresnel reflections at the front and back boundaries of the slab. Certainly such a configuration completely destroys invisibility. However, we have to make a distinction between the grating visibility and the slab visibility. Strictly speaking the grating visibility can be found by subtracting the reflection/transmission spectrum of the slab, $|R_0|^2/|T_0|^2$ from the reflection/transmission spectrum of the whole structure $|R_0+ R_1+ R_2|^2 / | T_0+ T_1+ T_2|^2$, i.e. $|R_0+ R_1+ R_2|^2-|R_0|^2$ or $|T_0+ T_1+ T_2|^2-|T_0|^2$. Therefore, the red curves in Fig. 6 (a) and Fig. 6 (b) provide the real indication of the *PT*-symmetric grating visibility from the right side. As we can from Fig. 6(b), the grating visibility is pretty low in reflection, particularly for $|\theta|<\theta_B|$.

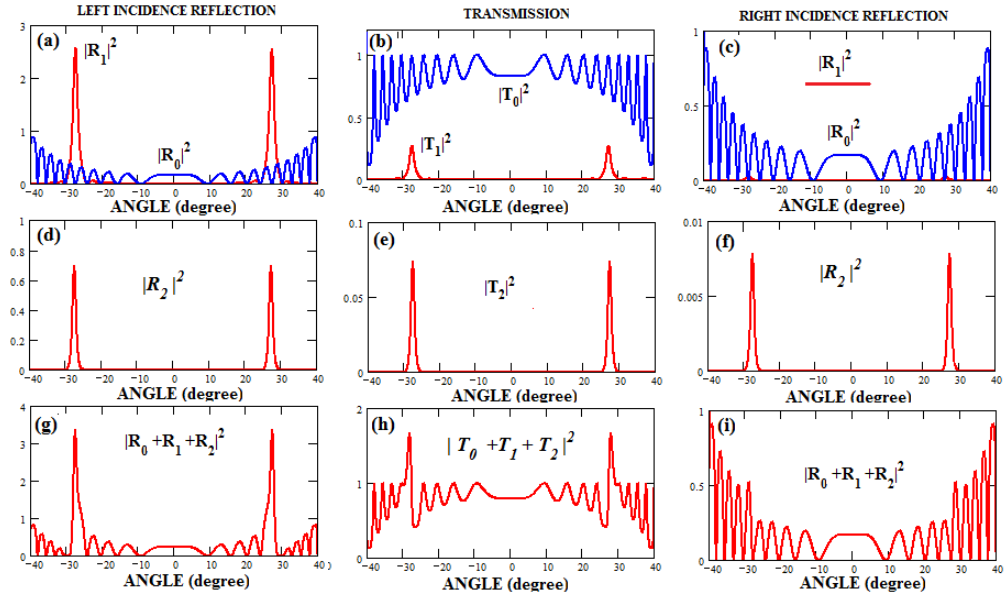

Fig. 5. The slab in air ( $\varepsilon_1 = \varepsilon_3 = 1$ $\varepsilon_2 = 2.4$ ). Left-side incidence: zeroth (blue), first order (red) (a), second (d) order reflection, as well as the combined coherent reflection (g) and zeroth (blue), first (red) (b) and second (d) order transmission, along with combined coherent transmission (h). Right side incidence: zeroth (blue), first (red) (c), second (f) order reflection and combined coherent reflection (i) as functions of the internal angle of incidence for $\Lambda=0.23$ μm. The other parameters are $d = 36\Lambda =8.28$ μm, $\lambda_0 = 0.633$ μm, $\xi=0.02$; $\theta_B=27.35^0$.

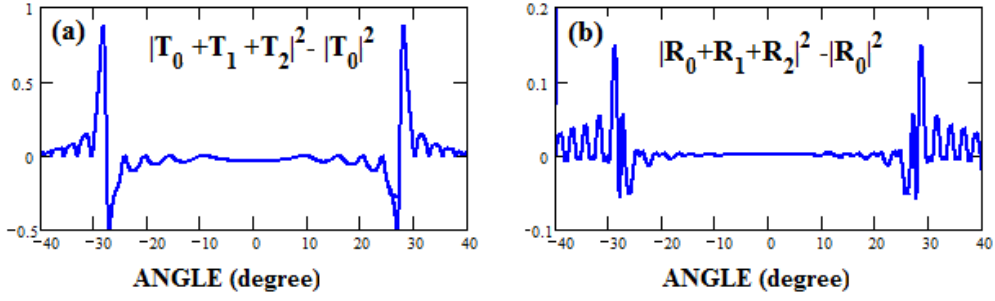

Fig. 6. The slab in air ( $\varepsilon_1 = \varepsilon_3 = 1$ $\varepsilon_2 = 2.4$ ): the grating visibility factor in transmission (a) and in reflection (b) for right-side incidence (from non-reflective side) as functions of the internal angle of incidence for $\Lambda=0.23$ μm (red, solid). The other parameters are $d = 36\Lambda = 8.28$ μm, $\lambda_0 = 0.633$ μm, $\xi=0.02$; $\theta_B=27.35^0$.

*4.2. Asymmetric slab configuration*

The grating can be attached to a substrate, and in fact such an approach is most practical for a grating which is a few microns thick. Possible configurations are presented in Fig. 7, where the non-reflective side of the grating is attached to the substrate (Fig. 7(a)) or where the reflective side is attached to the substrate (Fig. 7 (b)). In such a case of different media on the left and right sides, $\varepsilon_1 \neq \varepsilon_3$, the diffraction efficiency in transmission should be defined as $DET_m = \mathrm{Re}(\beta_0/\alpha_0)|T_m|^2 = (\varepsilon_3/\varepsilon_1)^{1/2}|T_m|^2$ [16], and the coherent diffraction efficiency is defined as $DET_{0+1+2} = (\varepsilon_3/\varepsilon_1)^{1/2}|T_0+T_1+T_2|^2$. It is easily shown that these are the same for left and right incidence for the same internal angle $\theta$.

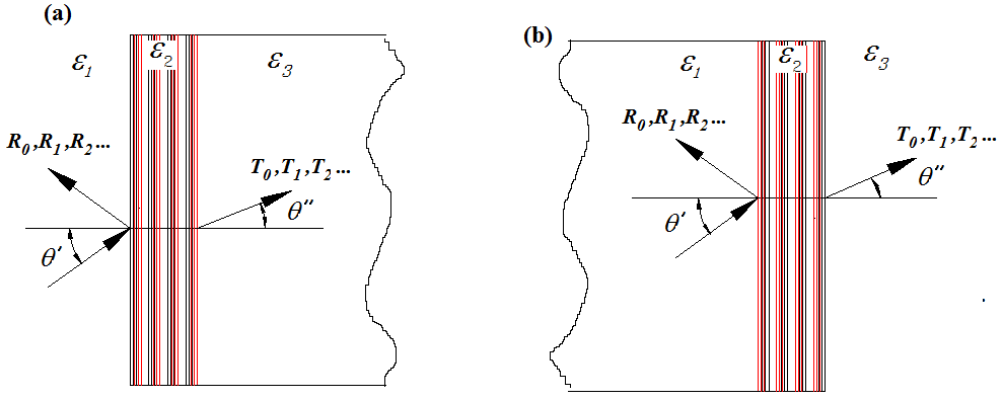

Fig. 7. Non-symmetrical configurations of the *PT*-symmetric grating on a substrate.

*4.2.1 Grating attached to left of substrate:* $\varepsilon_1 = 1; \varepsilon_3 = 2; \varepsilon_2 = 2.4$

Introduction of the substrate reduces the index contrast and as a result it reduces Fresnel reflection from rear boundary of the slab. In turn, it reduces the interference between zeroth order light in reflection and transmission, which is the main source of the invisibility distortion. However, regarding the visibility of the grating itself, we do not see any significant improvement in transmission Fig. 8 (h) and reflection Fig. 8 (i) compared with Fig. 5(h) and

Fig. 5 (i). This fact can be seen more clearly by comparing the grating visibility in reflection and transmission, i.e. comparing Figs. 6(a) and 6(b) with Figs. 9(a) and 9(b).

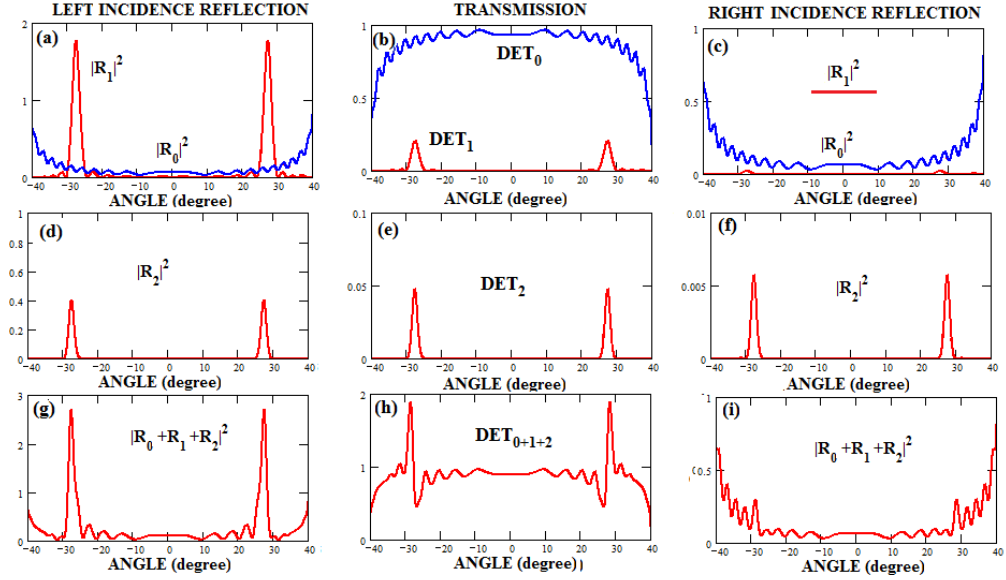

Fig. 8. The slab attached to the left of the substrate: ($\varepsilon_1 = 1$ $\varepsilon_2 = 2.4$ $\varepsilon_3 = 2.0$): Light incident from air side: zeroth (blue), first (red) (a) and second (d) order reflection, as well as the combined coherent reflection (g), and zeroth (blue), first (red) (b) and second (e) order transmission, along with the combined coherent transmission (h). Light incident from the substrate side: zeroth (blue), first (red) (c) and second (f) order reflection, along with the combined coherent reflection (i) as functions of the internal angle of incidence for $\Lambda=0.23$ μm. The other parameters are $d = 8.28$ μm, $\lambda_0 = 0.633$ μm, $\xi=0.02$.

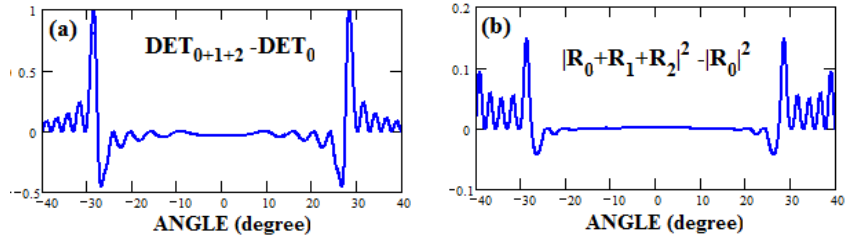

Fig. 9. The grating visibility when it is attached to the left of the substrate ($\varepsilon_1 = 1; \varepsilon_3 = 2$ $\varepsilon_2 = 2.4$) in transmission (a) and reflection (b) for right-side incidence (from non-reflective side). The other parameters are $d = 36\Lambda = 8.28$ μm, $\lambda_0 = 0.633$ μm, $\xi=0.02$; $\theta_B=27.35^0$.

4.2.2 *Grating attached to right of substrate*: $\varepsilon_1 = 2; \varepsilon_3 = 1; \varepsilon_2 = 2.4$

This configuration, shown in Fig. 7 (b), reduces the reflection from the front surface of the slab. As we can see from Fig. 10 (h) and Fig. 10 (i), the grating visibility is reduced to a very low level in transmission, and practically to zero (less than 1% diffraction efficiency) in reflection for light incident from the right. At the same time, reflection from the left side remains very strong (Fig. 10 (g)), with the main contribution coming from the first diffraction order (Fig. 10 (a) red curve). The significant reduction in the grating unidirectional visibility can be best seen by comparing its visibility in transmission (Fig. 11(a)) and reflection Fig. 11(b) with the same characteristics as the previous case, Fig. 9(a) and Fig. 9(b) respectively.

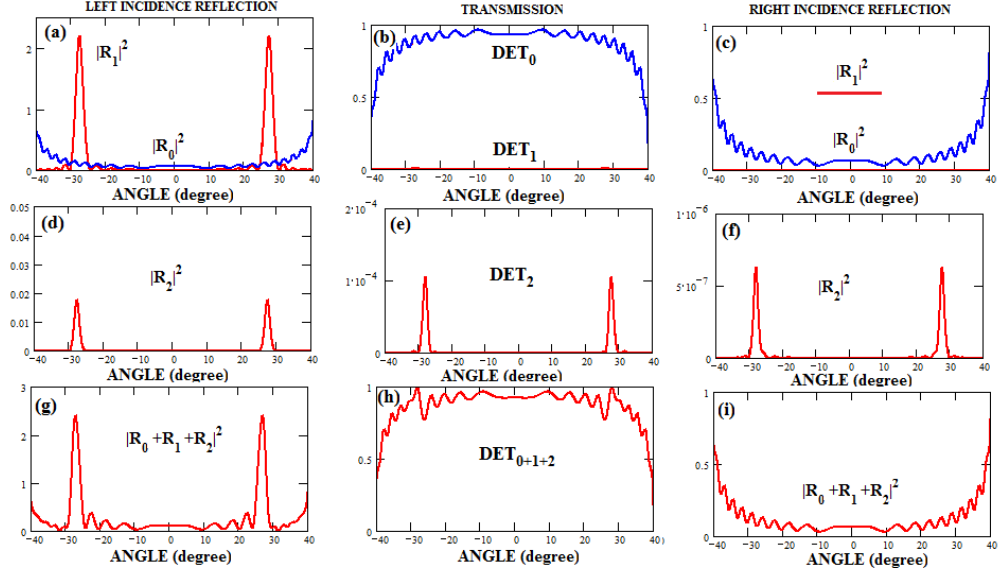

Fig. 10. The slab attached to the right of the substrate ($\varepsilon_1 = 2.0$  $\varepsilon_2 = 2.4$  $\varepsilon_3 = 1.0$). Light incident from substrate side: zeroth (blue), first (red) (a) and second (d) order reflection, as well as the combined coherent reflection (g), and zeroth (blue) and first (red) (b) and second (e) order along with the combined coherent transmission (h). Light incident from the air side: zeroth (blue), first (red) (c) and second (f) order reflection, along with the combined coherent reflection (i) as functions of the internal angle of incidence for $\Lambda$=0.23 μm. The other parameters are $d$ = 8.28 μm, $\lambda_0$ = 0.633 μm, $\xi$=0.02.

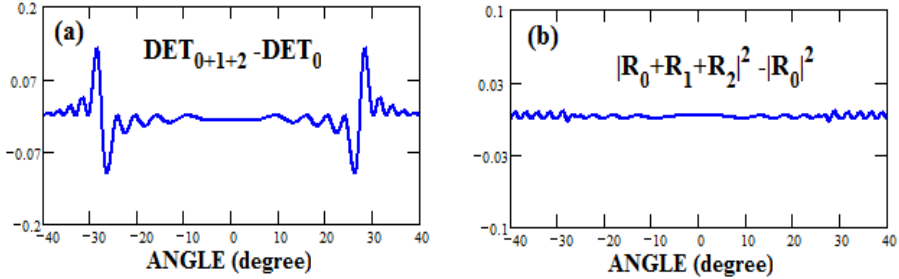

Fig. 11. The grating visibility when it is attached to the right of the substrate ($\varepsilon_1 = 2; \varepsilon_3 = 1$  $\varepsilon_2 = 2.4$) in transmission (a) and reflection (b) for right-side incidence (from non-reflective side). The other parameters are $d$ = 36$\Lambda$ =8.28 μm, $\lambda_0$ = 0.633 μm, $\xi$=0.02; $\theta_B$=27.35$^0$.

## 5.  Conclusions

Typically, diffraction characteristics are invariant for the direction of input light incidence. This empirical rule is generally known as Friedel's law. However, violations of this rule are observed in diffraction experiments and predicted by theoretical simulations in optics and atomic physics when diffraction occurs in *PT*-symmetric gratings. Such gratings are a combination of an even modulation of the index and an odd modulation of loss/gain in a media. The diffraction asymmetry reaches its highest level when the amplitudes of the modulation are equal. Such a balanced *PT*-symmetric grating diffracts light incident only from one side and might be completely invisible for light incident from the opposite side. The diffraction symmetry is lost because the index modulation has an inbuilt direction (it is symmetric under *PT*, but not under *P* itself).

In this paper we have studied such a *PT*-symmetric grating in a purely reflective mode when the fringes are parallel to the slab on which the grating is recorded. In such a grating geometry all diffraction orders starting from the first one contribute mostly near the first Bragg diffraction angle when an input light wave is incident from the reflective side of the grating.

The unidirectional nature of light interaction in the balanced *PT*-symmetric grating results in a simplified system of the rigorous coupled wave equations, which can now be solved analytically. In fact, there is no need to introduce any approximations like paraxiality, or diffraction only near the resonance, where only two diffractive modes are considered in order to bring the solution into a closed form. Preserving the second-order derivatives of the field amplitudes in the equations allowed us to keep all diffractive orders propagating in the forward and backward directions and to include boundary effects in the closed-form solution.

We have shown that the boundaries of the grating slab play a significant role in the total diffraction pattern. In addition to interference of non-diffracted (zeroth order) light reflected from the front and back boundaries of the slab, this reflected light experiences "secondary" diffraction by the *PT*-symmetric grating. For incidence from the left (reflective) side, Fresnel reflection from the slab boundaries produces significant diffraction in transmission in addition to the expected strong reflective diffraction.

Similarly, the Fresnel reflection strongly affects the unidirectional invisibility of the *PT*-symmetric grating for light incident from the non-reflective side of the grating. We have provided a detailed analysis of different configurations, such as filled space (no Fresnel reflection), the grating slab in air, and the grating slab attached to a transparent substrate with illumination from the air side and from the substrate side. It has been shown that Fresnel reflection from the front slab boundary (reflective side of the grating) produces much greater distortion of the otherwise perfect unidirectional invisibility.

The model developed, its closed form solution and the observations of this paper will be valuable for the accurate analysis of light diffraction on purely reflective *PT*-symmetric gratings, facilitating the investigation and design of novel devices in non-Hermitian optics.